\documentclass[doublecol]{epl2} 

\usepackage{amsmath,amsfonts}

\def\bra#1{\mbox{\boldmath $#1$}^{\top}}
\def\ket#1{\mbox{\boldmath $#1$}}
\newcommand{\bracket}[1]{\left\langle #1 \right\rangle}

\title{Detectability of the spectral method for sparse graph partitioning}
\shorttitle{Detectability of the spectral method for sparse graph partitioning} 

\author{T. Kawamoto\inst{1} \and Y. Kabashima\inst{1}}
\shortauthor{T. Kawamoto \etal}

\institute{                    
  \inst{1} Department of Computational Intelligence and Systems Science,
Tokyo Institute of Technology, 4259-G5-22, Nagatsuta-cho, Midori-ku, Yokohama, Kanagawa 226-8502, Japan
}

\abstract{
We show that modularity maximization with the resolution parameter offers a unifying framework of graph partitioning.
In this framework, we demonstrate that the spectral method exhibits universal detectability, irrespective of the value of the resolution parameter, as long as the graph is partitioned.
Furthermore, we show that when the resolution parameter is sufficiently small, a first-order phase transition occurs, resulting in the graph being unpartitioned.
}

\begin{document}

\maketitle

\section{Introduction}
Graph partitioning is often analyzed as a fundamental problem to understand the performance of community detection in complex networks.
Graph partitioning was originally an optimization problem: for a given number of modules, the problem is to find the partition with the sparsest cut under the constraint that the size of the modules are exactly or nearly equal.
When graph partitioning is applied to a social network with a modular structure, for instance, the nodes identified as members of a module are expected to belong to the same social group.

To clarify the perspective of graph partitioning as community detection, let us consider the partitioning of a uniform random graph (i.e., a random graph without planted block structures) as an example.
While it is still meaningful to find the optimal partition for each instance when the problem is regarded purely as an optimization problem, the result is hardly significant for community detection; it should be statistically different from the results for uniform random graphs.
Therefore, it is of significant importance to ascertain each algorithm's performance.
Interestingly, even when we generate random graphs with a planted modular structure, which have higher edge density within a module than between modules, the average performance of a partition may be indistinguishable from that of uniform random graphs as long as the modular structure is not sufficiently clear.
This indistinguishable region is called the \textit{undetectable phase}, while the region where the partition is positively correlated to the planted modules is called the \textit{detectable phase}. The boundary is called the detectability threshold \cite{Reichardt2008,Nadakuditi2012,Decelle2011a,Krzakala2013,Peixoto2013,Radicchi2013,ChenHero2015,Hu2012}.
Since many real networks are sparse, we focus on the case of sparse graphs. That is, the average degree does not increase as the total number of nodes increases.

Because the graph partitioning is usually formulated as a discrete optimization problem, which is computationally expensive, the spectral method \cite{Luxburg2007,Newman2006} that solves for the continuous relaxation of the original problem is often used.
Whereas the performance of partition generally depends on the choice of objective function to be optimized, it was shown in \cite{Newman2013} that the spectral method for three popular objective functions, namely modularity, normalized cut, and log-likelihood of the degree-corrected stochastic block model \cite{Karrer2011}, reduce to an eigenvalue problem of the normalized Laplacian when an elliptical normalization is considered as a constraint.

In this paper, we first show that the above three objective functions can be formulated in the framework of modularity maximization in the level of discrete optimization.
While the modularity form of the log-likelihood of the degree-corrected stochastic block model was already derived in \cite{Newman2013}, we show that the normalized cut can also be formulated in the same framework.
We then conduct a detectability analysis of the spectral method of the modularity with the spherical normalization constraint, i.e., the method with the modularity matrix.
The detectability analysis of the spectral method with the normalized Laplacian for sparse graphs was performed in \cite{Kawamoto2015Laplacian}.

An important difference between graph partitioning and community detection lies in whether the number of modules is given or to be estimated.
While the graph is always partitioned into a given number of modules when the normalized Laplacian is used, the method with the modularity matrix has its own criterion that determines whether the graph should or should not be partitioned.
We refer to the parameter region, where the graph is not partitioned, as the \textit{unpartitioned phase}.
To our knowledge, the detectability analysis was not concerned with this unpartitioned phase.
Focusing on the bisection problem, our analysis here reveals the relation between the detectable phase, undetectable phase, and unpartitioned phase.

\section{Unifying framework of graph partitioning}
We consider the bipartitioning of a graph $G(V,E)$ with a node set $V$ and an edge set $E$.
We denote the number of nodes as $N (= |V|)$ and the total degree, or the graph's volume, as $K (= 2|E|)$.
The two subsets of nodes obtained by the partition and their total degree are denoted by $S_{r}$ and $K_{r}$ ($r =1,2$), respectively.
We indicate the set of edges that connect nodes in $S_{1}$ and $S_{2}$ as $E(S_{1},S_{2})$.

The modularity with the resolution parameter, the objective function to be maximized, is
\begin{align}
Q_{\theta}(S_{1},S_{2}) = \sum_{r} \sum_{i,j \in S_{r}} \left( A_{ij} - \theta \frac{c_{i} c_{j}}{K} \right), \label{Modularity-1}
\end{align}
where $A$ is the adjacency matrix and $c_{i}$ represents the degree of node $i$.
The modularity function distinguishes the connectivity of the actual graph, $A_{ij}$, and the corresponding value of its null model, $c_{i} c_{j}/K$, in each module.
The resolution parameter $\theta > 0$ controls the balance between them.
Although the choice of the null model is generally arbitrary, as considered in most of the literature, we employ a random graph, the expected degree sequence of which is equal to that of the actual graph.
Because we focus upon bisection in the present study, the modularity function $Q_{\theta}$ can be expressed using a spin representation $s_{i} = \pm 1$ ($i = 1, \dots, N$).
Ignoring the constants irrelevant to the partition, we have
\begin{align}
Q_{\theta}(\ket{s}) = \bra{s}B\ket{s} = \bra{s} \left( A - \theta \frac{\ket{c} \bra{c}}{K} \right) \ket{s}, \label{Modularity-2}
\end{align}
where the vector $\ket{c} = (c_{1}, \dots, c_{N})$ has the degree of each node as its components and $\top$ indicates the transpose.
The matrix $B$ is called the modularity matrix, and $\theta$ is usually set to unity when bisection is considered.
As we show in what follows, this framework contains the normalized cut minimization as a special case.
The method of maximum log-likelihood is also a special case of this framework that has a particular value of $\theta$ \cite{Newman2013}.
A similar argument was presented in \cite{Delvenne2010}.

The objective function $f_{\mathrm{Ncut}}$ of the normalized cut is
\begin{align}
f_{\mathrm{Ncut}}(S_{1},S_{2}) = K \frac{|E(S_{1},S_{2})|}{K_{1}K_{2}} \label{Ncut-1}
\end{align}
for bisection.
We denote the minimum value of $f_{\mathrm{Ncut}}(S_{1},S_{2})$ as $\theta^{\ast}$, i.e.,
for any partition,
\begin{align}
& K \frac{|E(S_{1},S_{2})|}{K_{1}K_{2}} \ge \theta^{\ast}. \label{Ncut-2}
\end{align}
By using the relations $|E(S_{1},S_{2})| = \left(K - \bra{s} A \ket{s} \right)/4$, $K_{1} = (K + \bra{c}\ket{s})/2$, and $K_{2} = (K - \bra{c}\ket{s})/2$, we can recast (\ref{Ncut-2}) as
\begin{align}
\bra{s} A \ket{s} - \theta^{\ast} \frac{(\bra{c}\ket{s})^{2}}{K} \le K(1-\theta^{\ast}), \label{Ncut-3}
\end{align}
where we excluded the unpartitioned case, which is singular in (\ref{Ncut-2}).
Note that the right-hand side of (\ref{Ncut-3}) does not depend on the partition.
The equality holds when the left-hand side is maximized, which only occurs when the optimum partition is achieved unless nontrivial degeneracies exist; although the unpartitioned case also achieves the equality in (\ref{Ncut-3}), this choice is excluded. 
Therefore, if the optimum value $\theta^{\ast}$ is known, minimizing the normalized cut is equivalent to maximizing modularity with the resolution parameter, i.e.,
\begin{align}
\mathop{\mathrm{min}}_{\{S_{1},S_{2}\}} f_{\mathrm{Ncut}}(S_{1},S_{2}) &= \mathop{\mathrm{max}}_{\{S_{1},S_{2}\}} Q_{\theta^{\ast}}(S_{1},S_{2}). \label{ModularityFormNcut-2}
\end{align}
Because we do not know the minimum value of the normalized cut $\theta^{\ast}$ \textit{a priori}, the above argument is completely formal.
However, Eq.~(\ref{ModularityFormNcut-2}) denies the possibility that the optimum partition in the sense of the normalized cut may be different from the optimum partition in the sense of modularity.

We now consider the spectral method for (\ref{Modularity-2}).
As in \cite{Newman2006}, we relax the optimization of the spin variables $\ket{s}$ to a continuous vector $\ket{x}$ with the spherical normalization condition $|\ket{x}|^{2} = N$.
This leads to the eigenvalue problem of the modularity matrix $B$, and we determine the partition based on the signs of the leading eigenvector elements.
That is, each element in the leading eigenvector corresponds to the weight of a vertex, and we identify the set of vertices with the weights of the same sign as a module.
The unpartitioned phase is the case in which every node has the same sign of weight.
Note that the $1$-vector, the vector in which all elements are equal to unity, is not an eigenvector when $\theta \ne 1$. However, as we show, the leading eigenvector is orthogonal to the $1$-vector in the detectable region.

\section{Largest eigenvalue of the modularity matrix}
We analyze the performance of the spectral method for an ensemble of random graphs with a planted $2$-block structure.
We denote the node sets of planted modules as $V_{1}$ and $V_{2}$, where $|V_{1}| = p_{1} N$ and $|V_{1}| = p_{2} N$ ($p_{2} = 1 - p_{1}$).
We impose a constraint that the number of edges between blocks are $\gamma N$.
The rest of the edges are placed randomly within each module so that every node satisfies a given degree distribution $\{b_{t}\}$, where $b_{t}$ represents the fraction of nodes with degree $c_{t}$.
The average degree is denoted by $\overline{c}$ $(= \sum_{t}b_{t}c_{t}$).
As we have $\gamma = \overline{c} p_{1}p_{2}$ for a uniform random graph, it is natural to consider $\Gamma = 1 - \gamma/\overline{c} p_{1}p_{2}$:
$\Gamma = 1$ when modules are completely disconnected, and $\Gamma = 0$ for a uniform random graph.

Our goal is to evaluate the ensemble averages of the leading eigenvalues and their eigenvector distributions as functions of $\theta$ and $\Gamma$.
This allows us to measure the correlation between the partition obtained by the spectral method $\{S_{1}, S_{2}\}$ and the planted partition $\{V_{1}, V_{2}\}$.
As in the previous works \cite{Kabashima2012,Kawamoto2015Laplacian}, the largest eigenvalue of the modularity matrix $B$ can be calculated as
\begin{align}
\lambda_{1} = \lim_{\beta \rightarrow \infty} \frac{2}{N\beta} \ln Z(\beta | B),
\end{align}
where
\begin{align}
Z(\beta | B) = \int d \ket{x} \, \mathrm{e}^{\frac{\beta}{2} \bra{x} B \ket{x}} \delta(|\ket{x}|^{2} - N).
\end{align}

For the average largest eigenvalue, the \textit{replica trick} yields
\begin{align}
\left[\lambda_{1}\right]_{B}
&= 2 \lim_{\beta \rightarrow \infty} \frac{1}{N\beta} \left[\ln Z(\beta | B)\right]_{B}\nonumber\\
&= 2 \lim_{\beta \rightarrow \infty} \lim_{n\rightarrow 0} \frac{1}{N\beta} \frac{\partial}{\partial n} \ln \left[Z^{n}(\beta | B)\right]_{B}, \label{lambda1-1}
\end{align}
where we denote the ensemble average over random graphs as $[\cdots]_{B}$.
We consider the limit $N\rightarrow\infty$ and evaluate (\ref{lambda1-1}) using the saddle-point method with the replica-symmetric ansatz.
After a calculation analogous to that in \cite{Kabashima2012,Kawamoto2015Laplacian}, we arrive at a saddle-point expression of $\left[\lambda_{1}\right]_{B}$ (see Appendix for the specific form of $\left[\lambda_{1}\right]_{B}$).
It is composed of distributions of the order parameter functions $q_{r}(A, H)$ ($r=1,2$), which appear in $\left[Z^{n}(\beta | B)\right]_{B}$; their conjugate distributions $\hat{q}_{r}(\hat{A}, \hat{H})$; and auxiliary variables $\phi$ and $\hat{\Omega}$, which originate from the normalization constraint and the penalty term $(\bra{c}\ket{x})^{2}/K$.

Solving for the saddle point in the entire function space is, however, not feasible analytically or numerically.
Thus, we restrict the possibility of distributions $q_{r}(A, H)$ and $\hat{q}_{r}(\hat{A}, \hat{H})$ to simple forms of $q(A) = \delta(A - a)$ and $\hat{q}(\hat{A}) = \delta(\hat{A} - \hat{a})$.
While such distributions actually provide the exact saddle point for random regular graphs, they are approximations in general; this is called the \textit{effective medium approximation} (EMA).
Under this restriction, we can determine the average first eigenvalue $\left[\lambda_{1}\right]_{B}$ and the saddle-point conditions in analytic forms by using the functions
\begin{align}
& R_{n}(\phi, \hat{a}) = \sum_{t} \frac{b_{t} c_{t}^{n}}{\phi - c_{t}\hat{a}}, \\
& S_{n}(\phi, \hat{a}) = \sum_{t} \frac{b_{t} c_{t}^{n}}{(\phi - c_{t}\hat{a})^{2}},
\end{align}
and the moments $m_{nr} = \int dH \, q_{r}(H) H^{n}$ and $\hat{m}_{nr} = \int d\hat{H} \, \hat{q}_{r}(\hat{H}) \hat{H}^{n}$.
The average first eigenvalue $\left[\lambda_{1}\right]_{B}$ becomes
\begin{align}
\left[\lambda_{1}\right]_{B}
&= \mathrm{extr} \Biggl\{
\phi + 2 \hat{\Omega}\Omega - \frac{\theta}{\overline{c}} \Omega^{2} \nonumber\\
&\hspace{20pt} - \frac{\overline{c}}{a - \hat{a}} \left( \bracket{m_{2r}} + 2 \bracket{m_{1r} \hat{m}_{1r}} + \bracket{\hat{m}_{2r}} \right) \nonumber\\
&\hspace{20pt} + \frac{\overline{c}}{a^{2}-1}\left( a\bracket{m_{2r}} + \bracket{m_{1r}^{2}} \right) \nonumber\\
&\hspace{20pt} - \frac{\gamma}{a^{2}-1} \left( m_{11} - m_{12} \right)^{2} \nonumber\\
&\hspace{20pt} + R_{2}(\phi, \hat{a}) \left( \hat{\Omega}^{2} - 2 \hat{\Omega} \bracket{\hat{m}_{1r}} + \bracket{\hat{m}_{1r}^{2}} \right) \nonumber\\
&\hspace{20pt} + R_{1}(\phi, \hat{a}) \left( \bracket{\hat{m}_{2r}} - \bracket{\hat{m}_{1r}^{2}} \right)
 \Biggr\}, \label{lambda1-3}
\end{align}
where $\bracket{X_{nr}} = \sum_{r} p_{r} X_{nr}$.
The extremum conditions elucidate the appearance of each phase.
While we have $\hat{m}_{11} \ne \hat{m}_{12}$ with $\bracket{\hat{m}_{1r}} = 0$ in the detectable phase, the condition $\hat{m}_{11}^{2} = \hat{m}_{12}^{2} = 0$ is satisfied in the undetectable phase. The transition  occurs when $\phi$ and $\hat{a}$ satisfy
\begin{align}
S_{2}(\phi, \hat{a}) &= \frac{\overline{c} (1+\hat{a}^{2})}{(1-\hat{a}^{2})^{2}}. \label{GeneralDetectabilityThreshold}
\end{align}
In the detectable phase, $\phi$ and $\hat{a}$ are determined using the following extremum conditions:
\begin{align}
R_{1}(\phi, \hat{a}) &= \frac{\overline{c}\hat{a}}{1 - \hat{a}^{2}}, \label{phi-ahat}\\
R_{2}(\phi, \hat{a}) &= \frac{\overline{c}}{1 - \hat{a}^{2}} \left(\hat{a} + \frac{1}{\Gamma} \right), \label{nonequal-m1r}
\end{align}
while they are constant in the undetectable phase.
In both phases, we have $\hat{\Omega} = 0$, and the average first eigenvalue is
\begin{align}
\left[\lambda_{1}\right]_{B} = \phi.
\end{align}
Note that (\ref{GeneralDetectabilityThreshold}) does not contain the resolution parameter $\theta$; therefore, the detectability threshold is universal with respect to $\theta$.

There also exists a solution with $\hat{\Omega} \ne 0$ and $\hat{m}_{11} = \hat{m}_{12} \ne 0$.
This solution indicates the unpartitioned phase, and it is observed when the corresponding first eigenvalue becomes larger than that of the detectable and undetectable phases.
In this phase, they are determined using (\ref{phi-ahat}) and
\begin{align}
R_{2}(\phi, \hat{a}) &= \frac{\overline{c}}{1- \theta - \hat{a}}. \label{nonzero-hatOmega}
\end{align}
The transition to the unpartitioned phase occurs when the values of $\phi$ and $\hat{a}$ for two phases coincide.
The average first eigenvalue $\left[\lambda_{1}\right]_{B}$ does not have $\Gamma$-dependence and is constant in this phase.
As we observe in the examples below, while the transition from the detectable phase to the undetectable phase is continuous, the transition from the detectable phase to the unpartitioned phase is abrupt.

\begin{figure*}[t]
\centering
\includegraphics[width=2 \columnwidth]{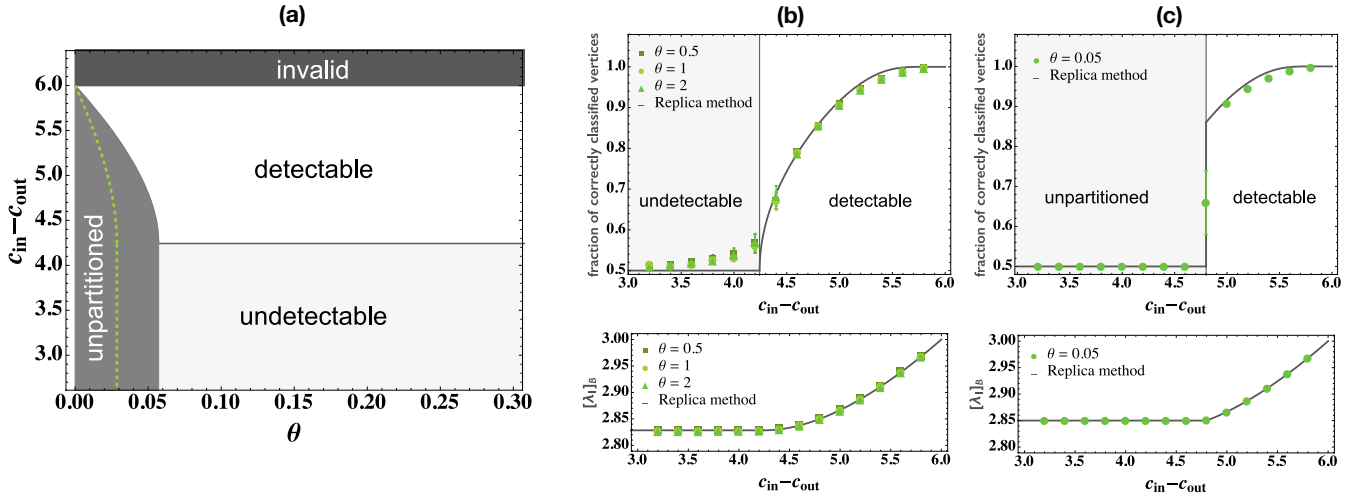}
\caption{
(Color online)
Left: (a) Phase diagram of the random 3-regular graph with respect to the resolution parameter $\theta$ and the strength of the block structure $c_{\mathrm{in}} - c_{\mathrm{out}}$.
The region $c_{\mathrm{in}} - c_{\mathrm{out}} > 2 c$ is invalid because we would have $c_{\mathrm{out}} < 0$.
Right: Fractions of correctly classified vertices and the eigenvalues of the modularity matrix in the random 3-regular graphs for (b) $\theta=0.5, 1, 2$ and (c) $0.05$. The solid lines are the analytical results. The dots are the results of the numerical experiments, where we set $N=10^{4}$, $p_{1} = p_{2} = 0.5$, and the average was taken over $20$ samples.
}
\label{RegularRandomDetectability}
\end{figure*}

\section{Random regular graph}
In the case of random regular graphs, the above results are exact, and we can analytically solve for the physical quantities and the boundaries of the phases.
The detectable phase of random $c$-regular graph has
\begin{align}
\left[\lambda_{1}\right]_{B} &= (c-1) \Gamma + \frac{1}{\Gamma}, \\
\hat{m}_{11}^{2} &= \frac{p_{2}}{c p_{1}} \left( 1 - \frac{1}{(c-1)^{2} \Gamma^{2}} \right) \left( (c-1)\Gamma^{2} - 1 \right).
\end{align}
The undetectable phase has
\begin{align}
\left[\lambda_{1}\right]_{B} &= 2 \sqrt{c-1},
\end{align}
and the detectability threshold is
\begin{align}
\Gamma = \frac{1}{\sqrt{c-1}}. \label{DetectabilityThreshold}
\end{align}
This is equal to the case of the normalized Laplacian \cite{Kawamoto2015Laplacian}.
Finally, the boundary of the detectable phase and the unpartitioned phase is
\begin{align}
\Gamma_{\mathrm{un}} &= \frac{c (1-\theta) + \sqrt{c^{2} (1-\theta)^{2} - 4 (c-1)}}{2 (c-1)}. \label{Regular-UnpartitionedThreshold}
\end{align}
This is a monotonically decreasing function that is minimum when
\begin{align}
\theta_{\mathrm{max}} = 1 - \frac{2 \sqrt{c-1}}{c}.
\end{align}
The corresponding value of $\Gamma$ coincides with the detectability threshold.
Note that when the graph is regular, the $1$-vector is the leading eigenvector in the unpartitioned phase.
Its eigenvalue is $c(1-\theta)$ and is equal to the eigenvalue of the undetectable phase at $\theta_{\mathrm{max}}$.
Hence, the region where both $\theta$ and $\Gamma$ are small is the unpartitioned phase.
Consequently, we obtain the phase diagram shown in Fig.~\ref{RegularRandomDetectability}(a).
Following the literature, we employed $c_{\mathrm{in}} - c_{\mathrm{out}}$, instead of $\Gamma$, to indicate the strength of the block structure;
in the case of equal-size blocks $p_{1} = p_{2} = 0.5$, $c_{\mathrm{in}} - c_{\mathrm{out}}$ is twice the difference between the average degree within a block and between blocks (see \cite{Kawamoto2015Laplacian} for the relation between $\gamma$ and $c_{\mathrm{in}} - c_{\mathrm{out}}$).

The fractions of correctly classified vertices and the average first eigenvalues $\left[\lambda_{1}\right]_{B}$ are plotted in Figs.~\ref{RegularRandomDetectability}(b) and \ref{RegularRandomDetectability}(c).
To draw the solid lines, we further approximated that the distribution of the eigenvector elements is Gaussian.
We can confirm a universal detectability curve for various values of sufficiently large $\theta$ and an abrupt transition between the detectable and unpartitioned phases for a small value of $\theta$.

Recall that in the case of the normalized cut, the resolution parameter $\theta$ is the optimum value of the objective function itself.
Although the exact value of $\theta$ is not known, it is bounded by the second-smallest eigenvalue of the normalized Laplacian \cite{Kawamoto2015Laplacian} using the Cheeger inequality \cite{chung1996spectral}.
The dashed line in Fig.~\ref{RegularRandomDetectability}(a) indicates the lower bound of $\theta$, i.e., one half of the second-smallest eigenvalue of the normalized Laplacian, while the upper bound is very large. 

\section{Stochastic block model}
\begin{figure*}[t]
\centering
\includegraphics[width=2 \columnwidth]{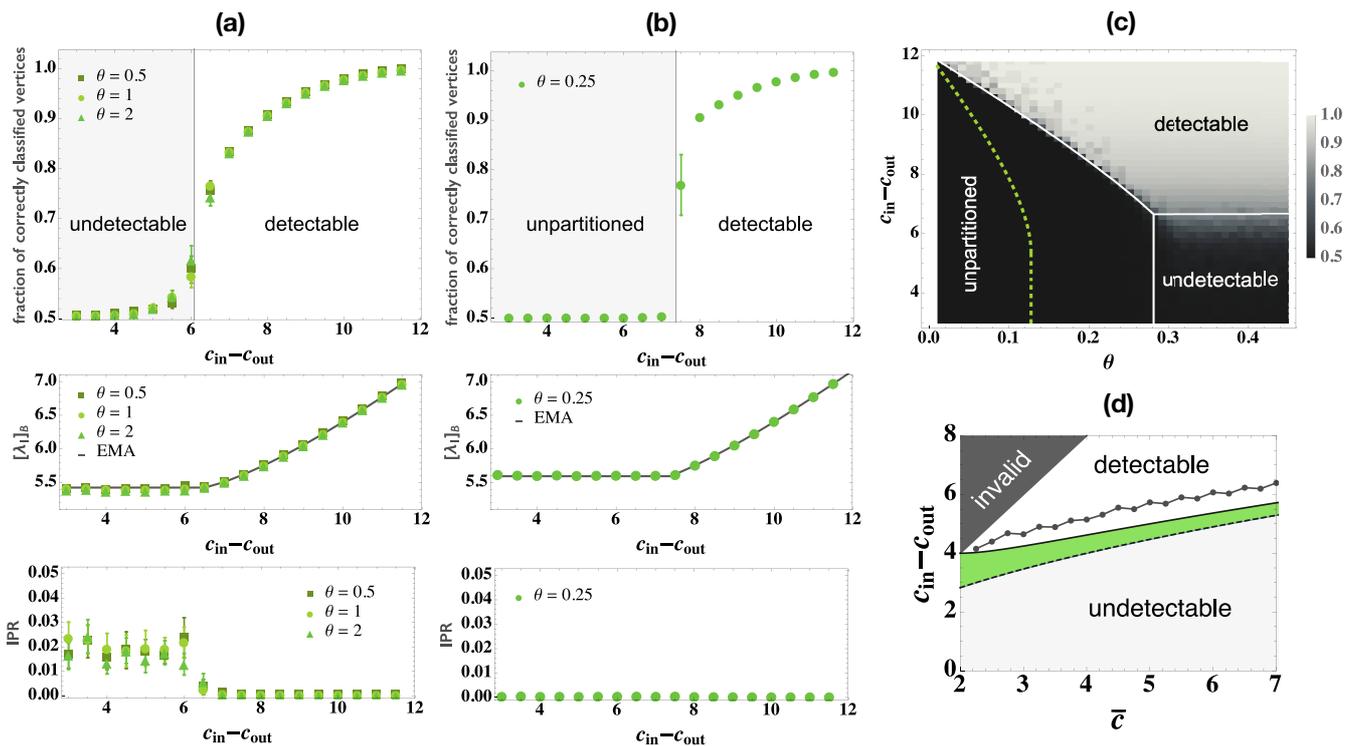}
\caption{
(Color online)
Behaviors of the spectral method for the stochastic block model with $\overline{c}=6$.
Left: Fractions of correctly classified vertices, average first eigenvalues $\left[\lambda_{1}\right]_{B}$, and IPR for (a) $\theta = 0.5, 1, 2$, and (b) $\theta = 0.25$.
The dots are the results of the numerical experiments, where we set $N=2\times10^{4}$, $p_{1} = p_{2} = 0.5$, and the average was taken over $20$ samples.
The solid line for $\left[\lambda_{1}\right]_{B}$ is the result of EMA.
Right: (c) Phase diagram of the fraction of correctly classified vertices with respect to the resolution parameter $\theta$ and the strength of the block structure $c_{\mathrm{in}} - c_{\mathrm{out}}$. 
Each point of the density plot indicates the average value over $20$ samples. 
The dashed line is the lower bound of $\theta$ estimated using the Cheeger inequality, where we used the EMA solution \cite{Kawamoto2015Laplacian} for the second-smallest eigenvalue of the normalized Laplacian.
The white lines indicate the results of EMA.
(d) Phase diagram of the detectability threshold with respect to the average degree $\overline{c}$.
The dashed line, solid line, and connected dots represent the estimates of the detectability thresholds with the dense approximation \cite{Nadakuditi2012}, threshold of the normalized Laplacian \cite{Kawamoto2015Laplacian}, and threshold of the modularity matrix with EMA.
Although they are not shown, we also confirmed the universal behavior with respect to $\theta$ for several unequal block sizes.
}
\label{ModularitySBM_cm=6}
\end{figure*}

Although there are many variants of the stochastic block model, we consider the most fundamental model.
Pairs of nodes within the same module and between different modules are connected with probabilities $p_{\mathrm{in}}$ and $p_{\mathrm{out}}$ ($p_{\mathrm{in}}>p_{\mathrm{out}}$), respectively.
That is, the nodes within the same module are more densely connected than the nodes in different modules.
Because we are focusing on sparse graphs, we set both $c_{\mathrm{in}} = p_{\mathrm{in}} N$ and $c_{\mathrm{out}} = p_{\mathrm{out}} N$ to be $O(1)$.
This model has the Poisson degree distribution, and therefore, our EMA-based treatment no longer offers the exact result.
Because no bound exists for the maximum of a node degree, we need to rely on the numerical estimate of the formal solution by truncating the infinite summations of $R_{n}(\phi, \hat{a})$ and $S_{n}(\phi, \hat{a})$.
The results of our replica analysis and those of the corresponding numerical experiments are compared in Figs.~\ref{ModularitySBM_cm=6}(a)--(c).
The comparison indicates that our estimates offer very accurate predictions.
The average first eigenvalue $\left[\lambda_{1}\right]_{B}$ is fairly large even in the undetectable phase, which is consistent with \cite{Guimera2004}.
Note that the degree of eigenvector localization, which we measured by the inverse participation ratio (IPR), increases gradually around the detectability threshold;
as far as we explored, the value of IPR becomes significantly large only in the undetectable phase (see \cite{Kawamoto2015Laplacian} for the definition of IPR).
Although it is difficult to prove whether the localized eigenvector is absent in the detectable region, its influence seems negligible.
This is consistent with what we observed empirically \cite{Kawamoto2015NBT}.

\section{Summary and discussion}
In summary, we showed that the modularity maximization with the resolution parameter $\theta$ offers a unifying framework of graph partitioning and analyzed the detectability of the spectral method of the unifying framework.
Our phase diagram shows that when the resolution parameter $\theta$ is sufficiently small, the unpartitioned phase appears before the detectability threshold because the block structure is weakened; that is, even when the block structure is statistically significant in the sense of the Bayesian inference, there exist cases where the graph has no significant structure in the sense of an objective function.
Otherwise, the detectability threshold is universal irrespective of $\theta$.
This behavior occurs probably because the graph is assumed to be infinitely large in our analysis;
because the order of the penalty factor $(\bra{c}\ket{x})^{2}/K$ is smaller than that of $\bra{x}A\ket{x}$ in the detectable phase, the value of $\theta$ does not affect the resulting performance.
Our results imply that, whereas the gap between the detectability threshold of the Bayesian inference and the spectral method was mainly due to the eigenvector localization when the normalized Laplacian was used \cite{Kawamoto2015Laplacian}, it is mainly because of the difference in the detectability threshold itself in the case of the modularity matrix.






\acknowledgments
The authors thank Takanori Maehara for helpful comments.
This work was supported by
JSPS KAKENHI No. 26011023 (TK) and No. 25120013 (YK) and
the JSPS Core-to-Core Program ``Non-equilibrium dynamics of soft matter and information.''

\section{Appendix: Derivation of the saddle-point expression}\label{ReplicaAnalysis}
From (\ref{lambda1-1}), the average of the largest eigenvalue $\left[\lambda_{1}\right]_{B}$ can be obtained by evaluating the moment $\left[Z^{n}(\beta | B)\right]_{B}$ of the ``partition function'' as an analytic function of $n$.
Assuming $n$ as a positive integer, we have
\begin{align}
&[Z^{n}(\beta | B)]_{B} \nonumber\\
&= \int \left(\prod_{a=1}^{n} d \ket{x}_{a} d\Omega_{a} \delta(|\ket{x}_{a}|^{2} - N) \, \delta\left( \Omega_{a} - \frac{\bra{c}\ket{x}_{a}}{N} \right) \right) \nonumber\\
&\times \exp\left( -\frac{\beta \theta N^{2}}{2K}\sum_{a=1}^{n} \Omega^{2} \right) \left[ \exp\left(\frac{\beta}{2} \sum_{a=1}^{n} \bra{x}_{a} A \ket{x}_{a} \right) \right]_{A}, \label{Z^n}\tag{A.1}
\end{align}
where $a$ is the replica index and the last factor represents the ensemble average over graphs with proper edge constraints.
We now introduce the following order parameter functions for the block $r$.
\begin{align}
\mathcal{Q}_{r}(\ket{\mu}) \equiv \frac{1}{p_{r}N} \sum_{i \in V_{r}} z_{i} \prod_{a=1}^{n} \delta(x_{ia} - \mu_{a}), \label{OrderParameterFunction}\tag{A.2}
\end{align}
where $z_{i}$ is an auxiliary variable that originates from the edge constraint.
Assuming that (\ref{OrderParameterFunction}) becomes invariant under any permutation of replica indices $a$ at the dominant saddle point, 
we then express these order-parameter functions and their conjugates as Gaussian mixtures as follows:
\begin{align}
&\mathcal{Q}_{r}(\ket{\mu}) = q^{0}_{r} \int dA dH \, q_{r}(A,H) \left( \frac{\beta A}{2\pi} \right)^{\frac{n}{2}} \nonumber\\
&\hspace{50pt}\times \exp\left[ -\frac{\beta A}{2} \sum_{a=1}^{n} \left( \mu_{a} - \frac{H}{A} \right)^{2} \right], \label{assumedQ} \tag{A.3}\\
&\hat{\mathcal{Q}}_{r}(\ket{\mu}) = \hat{q}^{0}_{r} \int d\hat{A} d\hat{H} \, \hat{q}_{r}(\hat{A},\hat{H}) \nonumber\\
&\hspace{50pt}\times \exp\left[ \frac{\beta}{2} \sum_{a=1}^{n} \left( \hat{A} \mu^{2}_{a} + 2 \hat{H} \mu_{a} \right) \right], \label{assumedhatQ} \tag{A.4}
\end{align}
where $q^{0}_{r} = \sqrt{(\overline{c} p_{r} - \gamma)/N p^{2}_{r}}$ and $q^{0}_{r}\hat{q}^{0}_{r} = \overline{c}$, as derived in \cite{Kawamoto2015Laplacian}.
These forms yield, in the limit $N\rightarrow\infty$,
\begin{align}
&\left[\lambda_{1}\right]_{B}
= 2 \, \mathop{\mathrm{extr}}_{q_{r}, \hat{q}_{r}, \phi, \Omega, \hat{\Omega}}
\Biggl\{
\frac{\phi}{2} + \hat{\Omega}\Omega - \frac{\theta}{2 \overline{c}}\Omega^{2} \nonumber\\
&\hspace{2pt} - \sum_{r} \frac{\overline{c}p_{r}}{2} \int dA dH dA^{\prime} dH^{\prime} \, q_{r}(A, H) \hat{q}_{r}(\hat{A}, \hat{H})
\left( \frac{(H + \hat{H})^{2}}{A - \hat{A}} - \frac{H^{2}}{A} \right) \nonumber\\
&\hspace{2pt} + \frac{1}{2N} \sum_{r,t}\sum_{i \in V_{r,t}} \int \prod_{g=1}^{c_{t}} \left( d\hat{A}_{g}d\hat{H}_{g} \hat{q}_{r}(\hat{A}_{g}, \hat{H}_{g}) \right)
\frac{\left(c_{t} \hat{\Omega} - \sum_{g=1}^{c_{t}}\hat{H}_{g}\right)^{2}}{\phi - \sum_{g=1}^{c_{t}}\hat{A}_{g}} \nonumber\\
&\hspace{2pt} + \frac{1}{4} \sum_{r,s} \left[ (\overline{c}p_{r} - \gamma)\delta_{rs} + \gamma (1-\delta_{rs}) \right] \int dA dH \int dA^{\prime} dH^{\prime} \nonumber\\
&\hspace{2pt}\times q_{r}(A,H) q_{s}(A^{\prime},H^{\prime}) \left( \frac{A^{\prime}H^{2} + 2HH^{\prime} + AH^{\prime 2}}{A A^{\prime} - 1} - \frac{H^{2}}{A} - \frac{H^{\prime 2}}{A^{\prime}} \right)
\Biggr\}. \label{lambda1-2} \tag{A.5}
\end{align}
In (\ref{lambda1-2}), $\hat{\Omega}$ is the conjugate of $\Omega$ in (\ref{Z^n}) and $\phi$ is another auxiliary variable that originates from the normalization constraint $|\ket{x}|^{2} = N$.

\bibliographystyle{eplbib}
\bibliography{refModDetectability}

\end{document}